\documentclass[journal,twoside,web]{ieeecolor}
\usepackage{tmi}
\usepackage{cite}
\usepackage{amsmath,amssymb,amsfonts}
\usepackage{algorithmic}
\usepackage{graphicx}
\usepackage{textcomp}
\usepackage{cite}
\usepackage{adjustbox}
\usepackage{dirtytalk}
\usepackage{mathtools}

\newcommand{\figref}[1]{Fig.~\ref{#1}}
\newcommand{\secref}[1]{Section~\ref{#1}}
\newcommand{\tabref}[1]{Table~\ref{#1}}

\newcounter{hypotcounter}
\newcommand{\hypotdef}[1]{%
\refstepcounter{hypotcounter}%
\textbf{H\thehypotcounter}%
\label{hypot:#1}
}

\newcommand{\hypotref}[1]{%
\textbf{H\ref{hypot:#1}}%
}

\usepackage{hyperref}

\def\BibTeX{{\rm B\kern-.05em{\sc i\kern-.025em b}\kern-.08em
    T\kern-.1667em\lower.7ex\hbox{E}\kern-.125emX}}
\begin{document}
\title{Zero-shot super-resolution with a physically-motivated downsampling kernel for endomicroscopy.}

\author{Agnieszka~Barbara~Szczotka,  Dzhoshkun~Ismail~Shakir, Matthew~J.~Clarkson, Stephen~P.~Pereira, Tom~Vercauteren
\thanks{
Manuscript received October 28, 2020. 
This work was supported by Wellcome Trust [203145Z/16/Z;203148/Z/16/Z], and EPSRC [NS/A000050/1;NS/A000049/1;NS/A000027/1]. This work was undertaken at UCL and UCLH, which receive a proportion of funding from the DoH NIHR UCLH BRC funding scheme. The PhD studentship of Agnieszka Barbara Szczotka is funded by Mauna Kea Technologies, Paris, France. Tom Vercauteren is supported by a Medtronic / Royal Academy of Engineering Research Chair [RCSRF1819/7/34].}
\thanks{A. B. Szczotka and M. J. Clarkson are with Wellcome / EPSRC Centre for Interventional and Surgical Sciences, University College London, London WC1E 6BT,
U.K.(e-mail:agnieszka.szczotka.15@ucl.ac.uk; m.clarkson@ucl.ac.uk}
\thanks{T. Vercauteren and D. I. Shakir are with  School of Biomedical Engineering \& Imaging Sciences, King's College London (e-mail: tom.vercauteren@kcl.ac.uk, dzhoshkun.shakir@gmail.com)}
\thanks{S. P. Pereira is with UCL Institute for Liver and Digestive Health, University College London, London
WC2R 2LS, U.K. (e-mail: stephen.pereira@ucl.ac.uk)}}

\maketitle

\begin{abstract}
Super-resolution (SR) methods have seen significant advances thanks to the development of convolutional neural networks (CNNs). CNNs have been successfully employed to improve the quality of endomicroscopy imaging. Yet, the inherent limitation of research on SR in endomicroscopy remains the lack of ground truth high-resolution (HR) images, commonly used for both supervised training and reference-based image quality assessment (IQA).
Therefore, alternative methods, such as unsupervised SR are being explored. 
To address the need for non-reference image quality improvement, we designed a novel zero-shot super-resolution (ZSSR) approach that relies only on the endomicroscopy data to be processed in a self-supervised manner without the need for ground-truth HR images.
We tailored the proposed pipeline to the idiosyncrasies of endomicroscopy by introducing both: a physically-motivated Voronoi downscaling kernel accounting for the endomicroscope's irregular fibre-based sampling pattern, and realistic noise patterns.
We also took advantage of video sequences to exploit a sequence of images for self-supervised zero-shot image quality improvement. We run ablation studies to assess our contribution in regards to the downscaling kernel and noise simulation. 
We validate our methodology on both synthetic and original data. Synthetic experiments were assessed with reference-based IQA, while our results for original images were evaluated in a user study conducted with both expert and non-expert observers. 
The results demonstrated superior performance in image quality of ZSSR reconstructions in comparison to the baseline method. The ZSSR is also competitive when compared to supervised single-image SR, especially being the preferred reconstruction technique by experts.
%Our proposed novel pipeline for improving quality of endomicroscopy in a self-supervised manner has the potential to replace the reconstruction method currently used in the clinic.

\end{abstract}

\begin{IEEEkeywords}
blind super-resolution, endomicroscopy,  video enhancement, zero-shot
\end{IEEEkeywords}

\section{Introduction}
\IEEEPARstart{P}{robe}-based confocal endomicroscopy (pCLE) is primarily used to assist in the diagnosis of cancer and several other conditions~\cite{Wallace2011}. The modality is based on the acquisition of video sequences and can be used to perform real-time optical biopsies in a clinical setup. Signal acquisition is through an optical fibre bundle and miniaturised optics, that can be integrated into endoscopic and needle-based devices for use in a wide range of interventional workflows.
%Needle-based CLE~(nCLE), a variant of pCLE, can for example be introduced via the working channel of standard instruments such as endoscopic ultrasound (EUS), leading to multi-modal EUS-nCLE devices~\cite{keane2019prospective}.
However, the very nature of the acquisition, via a bundle constructed of thousands of optical fibres, means that there is a significant limitation in the pCLE image quality. So, although the pCLE videos carry useful microscopic information, the clinical readability of the images remains hampered by the relatively low image quality.

The pCLE signals produced by each fibre are non-uniformly distributed over the
bundle Field of View (FoV) in a characteristic irregular and a quasi-hexagonal geometrical pattern, where each fibre is responsible for producing a single-pixel signal. The pixel signals contain both tissue signal and noise. Current pCLE reconstruction algorithms interpolate noisy pCLE signals onto an oversampled Cartesian grid. While this resampling allows compensating for artefacts such as the honeycomb pattern produced by non-Cartesian fibre arrangements, it does not have any denoising properties and introduces edge artefacts caused by linear interpolation across the edges of the underpinning Delaunay triangulation~\cite{tom_phd}.

There have been solutions developed for improving the pCLE signal to noise ratio and contrast, reducing triangulation artefacts~\cite{ravi2018effective,ravi2019adversarial} and dealing with the irregularity of pCLE signals~\cite{szczotka2020learning}. Recent solutions have mostly relied on the application of deep learning (DL) to pCLE images. More specifically, the majority of existing techniques for improving pCLE image quality have focused on single-image super-resolution (SISR) algorithms that are trained on pairs of low- and high-resolution (LR-HR) images. Unfortunately, in the context of pCLE, acquisition of HR videos is impossible, since high definition pCLE probes do not exist. The lack of HR pCLE thus severely limits the application of SISR to improve its image quality.
\par
Although attempts have been made to design SR techniques for pCLE, there is a clear need for a non-reference SR pipeline which does not need any ground truth images. Notably, there are still very few published works on blind SR for natural images ~\cite{shocher2018zero,ji2020real} and none for pCLE that discuss how to tackle SR on realistic images, as seen in clinical practice, without reliance on HR images.
\par
The influential work of Shocher et al.~\cite{shocher2018zero} gave rise to a renewed interest in using deep non-reference SR for natural images~\cite{ji2020real}. The core of the Zero-Shot (ZS) framework for SR relies on two things: an image to train a model in a self-supervised ZS manner, and a downscaling kernel to reduce the resolution of this image. In this paper, we expanded on the originally proposed ZSSR methodology~\cite{shocher2018zero} to maximise its benefits specifically for pCLE data. We propose a real-time ZSSR pipeline which uses a downscaling kernel, adapted explicitly for pCLE, and also a new ZS training strategy exploiting several primary frames from one pCLE video. 

The essential part of the framework is the downscaling kernel. The reconstructions obtained from models trained on the known realistic kernels had higher image quality than the images reconstructed from models trained with a bicubic kernel. Given the strong association between the downscaling kernel and the quality of the super-resolved image, it is essential to simulate LR pCLE as realistically as possible. To do so, we developed the necessary improvements by exploiting
the geometrical quasi-hexagonal fibre pattern and the nature of pCLE noise. We designed a downscaling procedure that encapsulates a more realistic Voronoi-based downsampling kernel and mimics realistic pCLE noise.
\par
Not only does ZS learning allow us to train a network without relying on HR ground truth images, but it also requires much less data to train the network. In the context of pCLE, we can nonetheless take advantage of the fact that the same bundle acquires all frames in any given
pCLE video. Under the assumption that the temporal non-linearity of the bundle transfer function is constant, we can assume that the frames in the video are correlated with each other by the same downscaling kernel. Thus, we can expand the training of our ZSSR to use 
a plurality of frames from the video instead of using only one frame from that video.

The main objective of this work is to investigate non-reference methods for improving the image quality of pCLE videos. This topic constitutes a new domain with largely unstudied potential, and to be best of our knowledge, and we are the first to explore it for pCLE. The main contributions of this work are listed below:
\begin{itemize}
    \item We design a Voronoi-based downscaling procedure which simulates realistic kernels that take under consideration the geometrical distribution of the pCLE signal sourced from irregularly distributed fibres.
    \item We propose to use several consecutive video frames during training to improve ZSSR in pCLE. Instead of using only a single image, we propose to take advantage of several consecutive frames from pCLE videos to extend our data set, allowing for more generalised ZS training.
    \item We demonstrate that the SR pipeline in conjunction with the training on data that realistically models the nature of the pCLE noise reduces the prominence of the imaging noise occurring in pCLE. 
 \end{itemize}
The result is the first real-time ZSSR pipeline using a downscaling kernel tailored to the physics of pCLE, and with enhanced denoising capabilities.

\section{Related work}
\subsection{pCLE HR images in the context of SR}
There has been extensive development in SR for natural images over last years~\cite{nasrollahi2014super}, and the state-of-the-art methods are in the majority based on DL~\cite{surveyWang}. Many studies have examined DL-based SR in medical imaging as well~\cite{LI2020}, with considerable interest in Generative Adversarial Networks (GAN) for image reconstruction and denoising~\cite{GAN_survey}. This section outlines the existing DL-based solutions presented in the literature for tackling specifically SR for pCLE with a focus on contending with the limitations in terms of availability of ground-truth HR data. The techniques for addressing this problem are focused on generating a substitute for HR pCLE.
\par
The authors of~\cite{shao2019fiber} built a custom optical system to acquire HR data for the supervised training of their DL reconstruction model. They used a dual-sensor system that allowed for capturing two images from the same light source by splitting the light beam in two: LR images acquired via a fibre bundle and HR images captured via an optical microscope. Such a system, due to its technical complexity, would not integrate well within the clinical workflow, would thus be unsuitable for large-scale representative data collection. 
\par
There have been some studies that have investigated the effectiveness of simulated data in the training of DL models.
In the first work on the application of DL to the pCLE SR task~\cite{ravi2018effective}, our team
simulated HR and LR pCLE images. HR pCLE images
were estimated with an offline registration and mosaicking method; synthetic LR images
were created from the HR images with the use of a physically-inspired simulation algorithm. This work has shown that DL models trained on such synthetic dataset translate well and can be used for improving the image quality of real pCLE images.
However, the proposed data synthesis suffers from several weak points of the registration such as limited availability of videos with undisturbed optical flow, lack of uniform super-resolution powers on the entire filed of view and misregistration artefacts.
\par
Rav\`i at al. use the potential of adversarial unsupervised training to get SR models without relying on HR pCLE images. Instead of pCLE data, natural images were used as a source of HR data for the training. To ensure that the GAN is trained towards generating meaningful pCLE-like reconstructions, this work added a physically-inspired cycle consistency loss based on the the reconstruction method using fibre pattern presented in~\cite{ravi2018effective}.
%Notably, there is a great deal of debate surrounding 
Even though many successful works on GAN have been publishing, the downside of that methods is that there is still uncertainty about whether GAN-based models generate results which deviate from the target domain. 
\par
Our previous work on how to input irregular pCLE signals into a CNN framework~\cite{szczotka2020learning} used the algorithm from~\cite{ravi2018effective} to simulate synthetic data. To make a systematic comparison between different types of CNN layers, we used histopathological images as the equivalent of HR images to simulate LR pCLE-like images. %This simulation procedure can be thought of as a variant of earlier ones such as those used in~\cite{ravi2018effective} since it shared the same reconstruction algorithm and select fibre patterns. %and serves as partly in design of our proposed downscaling procedure. 
The main drawback for pCLE is that this work is limited to the use of synthetic images based on histopathologies without consideration of the real pCLE images. 
\par
pCLE images are very similar to CLE images, and that has attracted some research efforts in developing SR techniques~\cite{izadi2019image,izadi}.
%This is the closest research to our pCLE SR exploration, yet very different in principle.
These two methods developed to improve CLE images address only synthetic evaluation of SR, without showing the adaptation to non-reference SR.
In both works, the authors used synthetic LR images, and train networks against original CLE images as HR images. They have shown excellent properties of their neural architectures being able to recover HR images from LR, yet both cases use a synthetic study where improvement for the original CLE images is not explored.
Moreover, their synthetic experiments are limited to using a bicubic kernel as the downscaling method to generate LR images from the CLE images. Yet, since the CLE relies on a different acquisition geometry than that in pCLE, this kernel is unsuitable for use in pCLE.

\subsection{Zero-shot SR}
Zero-shot (ZS) learning allows learning a data representation of samples
on the spot. 
From a minimal dataset
given for inference, ZS allows to train models capable of generalising to this data stemming from a distribution which was not seen during training time.
In the context of SR, the research work~\cite{shocher2018zero} explores information redundancy within one frame only to train an image-specific CNN network capable of predicting an super-resolved image from itself. It was shown that natural images contain recurring patches that capture the image representation well, so they may contribute to improving image content through the use of SR approaches~\cite{glasner2009super}.
Based on that finding, Shocher et al. used the LR frame as a source of LR-HR patches for training the image specific network to improve image resolution~\cite{shocher2018zero}. These patches are downscaled with either a bicubic kernel or an estimated kernel to simulate signal loss. Extracted patches are further augmented with eight rotation/flips to generate a bigger dataset. Pairs of LR-HR patches are used in a supervised manner to train a network. It is important to give attention to the fact that the input image has a dual role in ZSSR. 
It serves as the source HR image giving rise to the paired LR-HR image patches for training, and also serves as the LR image at inference time. Predicted SR estimations are added to the training data set for further gradual training. 

\subsection{Downscaling kernels for SR}
Previous studies on DL-based SR have primarily concentrated on using a default bicubic kernel with anti-aliasing as counterparts of real downscaling kernels. 
The NTIRE challenge~\cite{Timofte_2017_CVPR_Workshops,LI2020} aims at developing SR methods able to not only perform well on simulated images with a known bicubic kernel but also on \say{real} cases with unknown downscaling kernels. The results of the challenge demonstrated that the misestimation of the downscaling kernel affects the quality of super-resolved images. Furthermore, the authors of the original ZSSR work acknowledge these limitations, and they confirmed in their analysis that the performance of their SR pipeline is strongly affected by  the choice of downscaling kernel. Real images are influenced by limitations of the imaging system such as sensor noise, non-ideal point spread function, and reconstruction artefacts.
In reality, the downscaling kernel is typically not known, and when choosing one, there are several important technical considerations to be taken such as the optical model, the noise model etc.
Despite considerable efforts in the past, precise kernel estimation has proved a difficult goal. DL models will underperform when trained on pCLE images downscaled with default bicubic kernel, as it departs considerably from the acquisition physics of pCLE.
All these are downgrading factors and influence the acquisition model so should be considered when estimating a realistic downscaling kernel.

The problem of estimation of the downscaling kernels is not a new one.
In the pre-DL SR literature, Irani et al., for example, blindly estimated kernels and showed that it improves their super-resolution~\cite{michaeli2013nonparametric}.
In order to improve super-resolved images, researchers are currently looking at better ways to include downscaling kernels into non-reference SR pipelines. 
For instance, in \cite{shocher2018zero}, the authors have fused both previously developed classical kernel estimation techniques~\cite{michaeli2013nonparametric} with novel DL architecture. 
Although there are many classical solutions, the field of DL methods is evolving rapidly with much space for the development of novel deep learning methods with realistic kernels.

\subsection{Noise modelling in pCLE}
The works exploiting DL for SR are extensive, but they are primarily concerned with artificially generated LR images lacking any noise, and only recently the interest in real real-world image SR increased \cite{lugmayr2020ntire}. Practically, noisy images are typically considered for denoising tasks \cite{izadi2019whitenner}. Yet, there is a demonstrated improvement coming from using SR and denoising jointly \cite{zhang2017beyond,zhang2018learning, ji2020real}.
In~\cite{shocher2018zero}, the authors have shown that adding a small amount of Gaussian noise to LR images appears to have a positive impact on their ZSSR pipeline. They noticed that the optional addition of noise helps the network to distinguish between the correlated mapping of LR-HR pairs and uncorrelated noise. They showed that when models are trained with noise, it directly impacts SR results towards higher PSNR~\cite{shocher2018zero}. The pCLE noise contributes importantly to lower the pCLE image quality creating characteristic textured noise patterns
affecting many neighbouring pixels around a noisy fibre. Until now, there has not been any systematic study considering the impact of simulating noise in order to train an SR network for the additional task of denoising in pCLE. To the best of our knowledge, there is hardly any published evidence in the literature regarding the effects of using noisy LR pCLE in ZSSR.

\section{Material and Methods}
\subsection{Material}
\label{data}
In our experiments, we used two data sets: the first composed of clinical pCLE video sequences and the second being a simulated video endomicroscopy dataset. The clinical data is used to test our methodology qualitatively, and the simulated data are used to test our methodology quantitatively.

Although we can test the image quality (IQ) improvement qualitatively, this remains a complex and somewhat subjective task for human raters. We thus also use additional synthetic data to address the apparent lack of HR images for quantitative analysis. The pseudo endomicroscopies are a synthetic equivalent of real HR pCLE, and their availability facilitates quantitative validation of the proposed methodology. To obtain the pseudo endomicroscopy dataset we simulate it with the method originally introduced in~\cite{szczotka2020learning}. This method mimics both signal distribution through a geometrical position of the fibres in the bundle and generates realistic pCLE noise. Thanks to that, the resulting images are characterised by typical triangulation artefacts and noise patterns, making our synthetic test cases more similar to real pCLE than basic simulation and hence reduce the domain gap.

For simulating synthetic endomicroscopy, we utilise the histological dataset named Kather\footnote{\url{https://doi.org/10.5281/zenodo.1214456}} published in~\cite{kather2016multi}.
This dataset is a good choice for two reasons. First, histological images show structures similar to those which can be seen on real pCLE. In particular, the Kather dataset contains multi-class textures found in colorectal cancer histology. Secondly, the histological data were digitised with an HD camera and saved using lossless compression.
The images thus show tissues in very high-quality magnification without any significantly visible noise and compression artefacts. These high-quality images serve well as the equivalent of ground truth images.

We selected seven original pCLE video sequences from the pCLE-oriented education platform Cellvizio.net\footnote{Available at \url{http://www.cellvizio.net/}}. The content was selected ensuring that our methodology is tested for diverse types of tissue and cancer stages. Examples of the selected video exports with their SR versions can be found in the supplementary material.

\subsection{Zero-short super-resolution architecture and training}
\label{sec:zssr}
\begin{figure}
  \includegraphics[scale=0.35]{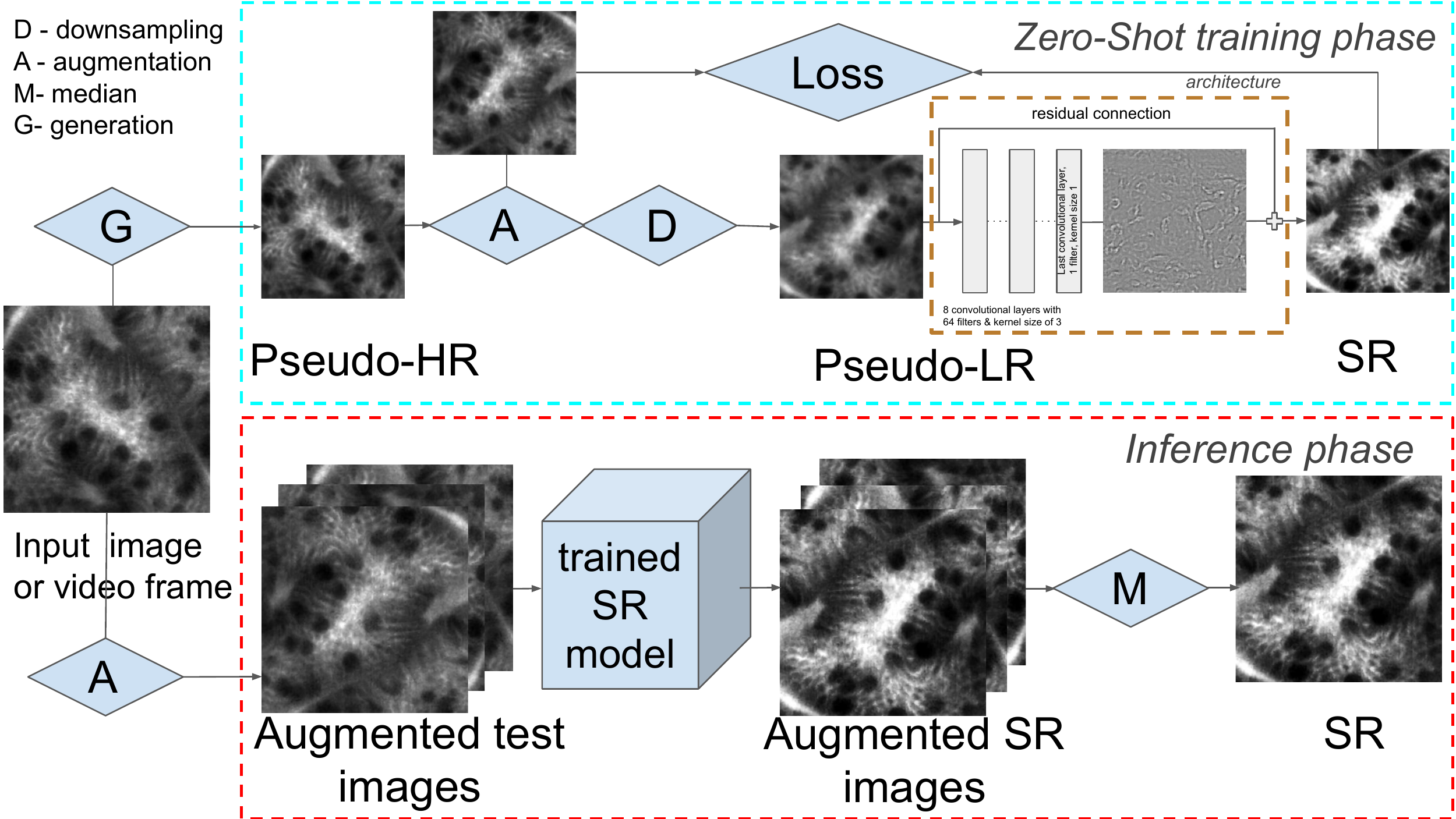}
    \caption{Graphical illustration of the proposed pCLE video ZSSR pipeline. The cyan (upper) block represents the training phase, and the red (lower) box represents the inference phase. The training starts from the input image or video. Note that, by design, the same input is used for both (zero-shot) training and inference. To extend the pipeline for handling multi-frames video, we iterate through a video stream, and each frame one by one becomes the input image. The input is used to generate~(\textbf{G}) the pseudo-HR, then the downscaling (\textbf{D}) creates the pseudo-LR with a Voronoi (or bicubic for comparison) kernel with augmentation (\textbf{A}) is also used for the training of the SR network. For the inference phase, the input video in its original size and augmented to create eight images per frame enhanced by a trained SR network. The final SR is a median (\textbf{M}) of these images after reverting the augmentation to the original frame. Detailed description is in~\secref{sec:zssr}.}
  \label{fig:pipe}
\end{figure}

We designed a ZSSR pipeline (cf. \figref{fig:pipe}) tailored to video pCLE taking inspiration from the original ZSSR paper~\cite{shocher2018zero}. Formally, HR image $I_{HR}$ is related to LR image $I_{LR}$ by degradation process $D$ defined as:
\begin{equation}
    I_{LR}=D({I}_{HR},\rho)
\end{equation}
\begin{equation}
    D({I}_{HR},\rho) = {I}_{HR}\downarrow_s + n_{\sigma}, \{s, \sigma\} \subset \rho,
\end{equation}

where $\downarrow_s$ is downscaling operation and $\rho$ are parameters of degradation process including but not limited to scaling factor $s$ and Gaussian noise $n$ with standard deviation $\sigma$.
The blind ZSSR aims in recovering HR approximation of the ground truth $\widehat{I}_{HR}$ from $I_{LR}$ under the assumption of known $D$:
\begin{equation}
  \widehat{I}_{HR} = F(I_{LR}; \theta, D({I}_{LR},\rho)),   
\end{equation}
where $F$ is the super-resolution process and $\theta$ denotes the parameters of $F$. In blind ZSSR $I_{HR}$ is not available. The optimisation of $F$ is based on the assumption that $F(\theta,D)$ is preserved with different scales and does not change depending on the input image scale. Thus, instead of $I_{HR}$, we used $I_{LR}$ to get its LR version $I_{LR}\downarrow$ based on $D$. As the results, the objective of ZSSR is to find $F$ by minimising the loss function $loss$ between the generated $\widehat{I}_{HR}$ and the test image ${I}_{LR}$. 
\begin{equation}
    \widehat\theta = \underset{\theta}{\arg\min}\;loss(
\widehat{I}_{HR}, I_{LR}) + \lambda\omega(\theta),
\end{equation}
where $\omega(\theta)$ is the regularisation term, and $\lambda$ is the trade-off parameter.

%We expected that an SR model trained on self-generated LR-HR paired image patches could recover the improved version of the input image.
%Unlike SISR, which generally needs a significant amount of data to be trained well, ZSSR framework enables reference-free SR with using only one image and downscaling kernel. Thanks to ZS-based training, we can improve the input image only with information coming from the input image itself.
%The trained model can then also be applied to new but similar images, as is the case for frames in a video sequence. 
%The pipeline is presented in \figref{fig:pipe}. 

\subsubsection*{Network architecture} Identically to~\cite{shocher2018zero}, we build an eight-layer CNN network with 64 kernels and ReLU as the activation function in each layer. We also use one skip-connection to learn the residual image instead of mapping it directly. We have shown that averaging of penultimate CNN layer~\cite{szczotka2020learning} improves SR, thus differently to~\cite{shocher2018zero}, in the proposed network the last CNN layer uses a kernel of size one and the linear activation function to reconstruct an SR image instead of learning another kernel.

\subsubsection*{Training procedure}
Our training approach is illustrated in \figref{fig:pipe}.
In brief, training of the network starts from the input image. First, the input image is transformed to a pseudo-HR image, and next downscaled to a pseudo-LR image, which is further augmented to generate a set of pseudo-LR images. The training of ZSSR, marked by the cyan dashed box in \figref{fig:pipe}, is supervised on the LR-HR image patches. The patches are extracted from the pseudo-HR pCLE. 
A detailed description of the bespoke training data generation is provided in \secref{downscaling}. In this paragraph, we focus on our adaptations of the training procedures presented in~\cite{shocher2018zero}.
\par
pCLE has a circular FoV, resulting in an all-black region outside the circular region of interest. To avoid performing convolutions with those non-informative black pixels, we cropped rectangular patches from the pseudo-HR and LR images. A patch size of 340 pixels was chosen as this is the largest crop region that can be applied across all videos. The central rectangular crop also facilitates augmentation, in particular rotation. We augmented the test images by using a combination of four rotations of $\{$0, 90, 180, 270$\}$ degrees and two flips \texttt{\{left, upside down\}}, yielding eight unique augmentations per image. 
\par
Experiments on pCLE videos exploit multiple frames, instead of using the only first frame from the sequence. We iterate through each video to extract several frames into the training set. These input train data are used during training of ZSSR video. Each image in the train set is cropped into the patch and augmented as we describe it for the input image in the previous paragraph.
\par
In the original ZSSR work, the pipeline benefits from tests performed during training.
Shocher at al.~\cite{shocher2018zero} evaluate their model several times during training and create intermediate predictions. They add them to the training data as \say{HR fathers} to enhance the training dataset. In a similar vein, we evaluate our model 10 times during training and every time we add the newly predicted image to the input data set, as the HR image.
After every 100 epochs, we use the test image to evaluate the current model. The super-resolved image produced on each evaluation becomes the new \say{HR father}, and it is added to the training set. Naturally, this grows the training set by one image each time, eventually leading to a ten-image training set in the end. To sum up, after performing 100 epochs, the model is evaluated. This is repeated until a total of 1000 epochs is reached.
In the original work, authors sample \say{HR fathers} from the training dataset with a lower confidence level than the input image augmentations to avoid learning the wrong representation. In contrast, we unified the sampling to be equal for all images in our training set to avoid tuning additional hyperparameter (sampling rate) and simplify the implementation of our pipeline.

We trained the network with Adam as the optimiser with  $\beta_1=0.9$, $\beta_2=0.999$.
We update the learning rate $lr$ periodically with an exponential decay
defined as:
%\begin{equation}
$
    %\text{decayed learning rate} = \text{learning rate}
    lr \leftarrow lr
    \times 
    %\text{decay rate}
    dr ^ {\frac{\text{global step}}{  \text{decay steps}}}
%\end{equation}
$
where empirically we set the decay rate $dr$ to 0.95 and decay step to 1000. Training starts with the learning rate of 0.001, gradually decreasing until it reaches $10^{-7}$ at the end of the training.
Additionally, we use mini-batch training with a batch size of 8 to update weights more reliably than when only one training image is used.

\subsubsection*{Choice of loss} 
Zhang et al. have shown that features extracted by any deep network can be adapted as a perceptual similarity metric~\cite{zhang2018perceptual}.
They designed a framework called LIPIPS to calibrate three deep networks with scores of the images judged by human raters. Their approach improves the performance of the network in terms of similarity to human perception and outperforms the widely used PSNR and SSIM metrics. 
Another observation of theirs is that any DL network, regardless of training style, architecture or data used for training its weights, is sufficient to play the role of the general feature extractor in this context. 
Based on those observations, we decided to utilise the power of their framework to our pCLE application.
% We believe that LIPIPS 

\begin{figure}%[!htb]
\centering
  
    \includegraphics[scale=0.25]{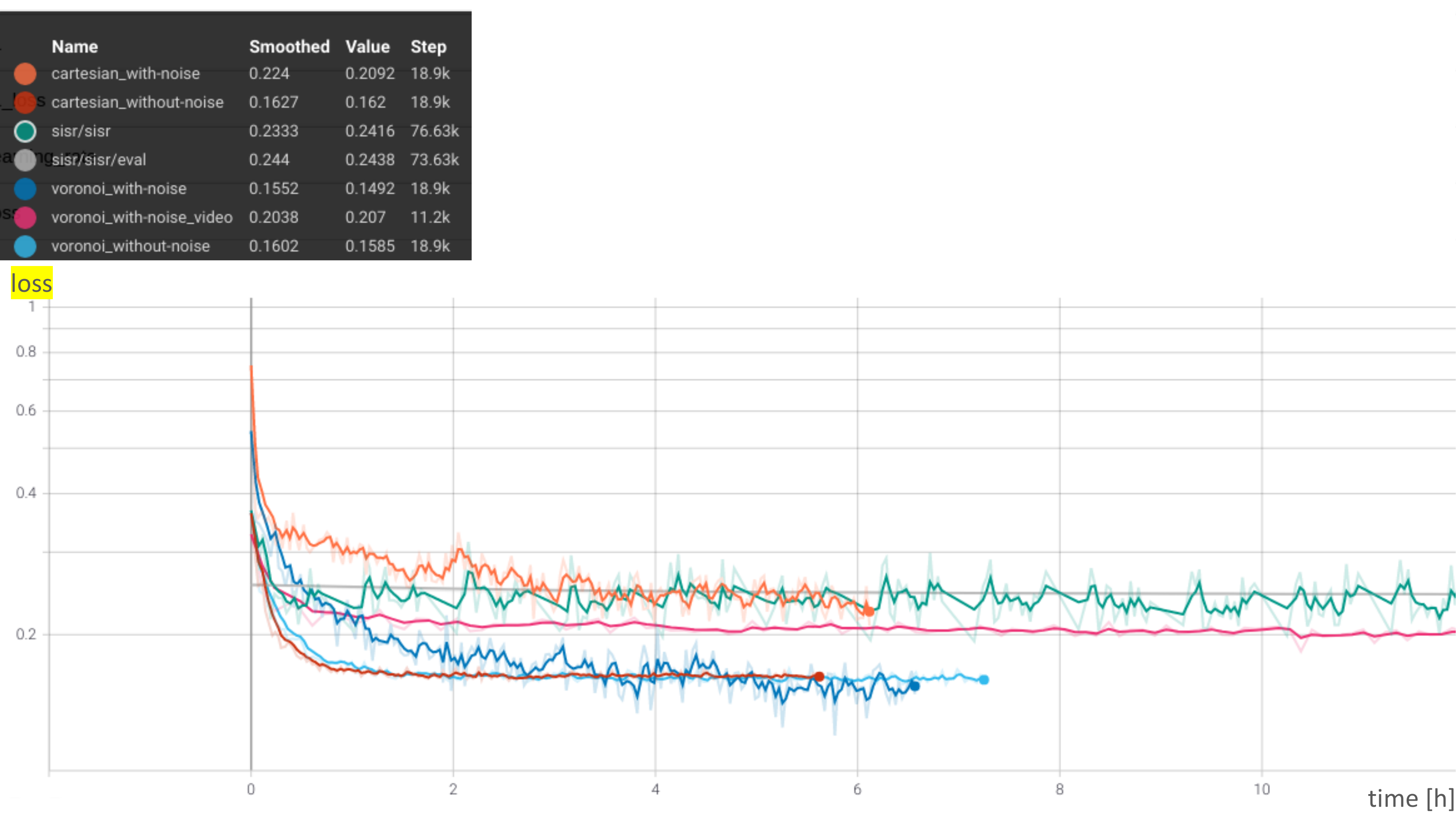}
\caption{Sample training losses}
\label{fig:loss}
\end{figure}

Following initial experiments with using the reference-based IQA metric LIPIPS~\cite{zhang2018perceptual} directly as a loss,
we built a custom loss function by introducing an additional L1-based term.
We used the pre-trained LIPIPS network\footnote{Available at \url{https://github.com/richzhang/PerceptualSimilarity}}, setting the linear calibration parameter to \texttt{net-lin} and selecting the VGG architecture as the network variant parameter. The network is pre-trained for RGB images (three colour channels). pCLE videos have only one channel, so we converted our videos to RGB by replicating the available scalar channel. Our initial experiments, 
%quickly discovered
confirmed
that LIPIPS works well, as expected, for generating sharp and detailed images. Yet it enhances all small details in the pCLE images, including the characteristic pCLE noise. We observe that LIPIPS is not robust enough to handle pCLE noise. To help our model distinguish noise from the pCLE signal, we regularise the loss by adding an L1 norm term. It steers the model in the direction of denoising. 
In summary, our new loss is defined as:
\begin{equation}
\begin{split}
    loss = &\underbrace{
    \sum_{l}\frac{1}{H_l,W_l}\sum_{w,l}\parallel w_{l}	\odot \hat{\vartheta}_{r}^{l} - \hat{\vartheta}_{p}^{l} \parallel_2^2 }_{\text{LIPIPS}} \\
    + & \underbrace{\lambda \frac{1}{n}\sum^{n}_{i=1}|I_{r}-I_{p}|}_{\text{L1}},
\end{split}
\end{equation}
where LIPIPS term~\cite{zhang2018perceptual} is computed with features given for a layer $l$ as $\hat{\vartheta}_{r}^{l}$, $\hat{\vartheta}_{p}^{l}$ $\in$ $\mathbb{R}^{H_l\times W_l\times C_l}$ that are extracted from a reference $I_{r}$ and a predicted $I_{p}$ image respectively. The LIPIPS term is a channel-wise $C$ sum of an average $L2$ distance in spatial dimension \{$W$, $H$\} of the feature stack scaled by vector $w^l\in \mathbb{R}^C_l$. $L1$ term is an average distance between reference and prediction scaled by $\lambda$, which was set empirically to 5. The sample training losses are in~\ref{fig:loss}.
\subsubsection*{Using the trained model for predictions}
The inference is marked by the red box in \figref{fig:pipe}. Predictions are generated from the trained model by inference from the video stream, which first frame or several frames were used for training depending on the experiment setup. For prediction, we use original images, as seen in the clinic. The generation of a pseudo-HR is only used for training purposes.
The near real-time inference per frame of our shallow network allows the predictor to efficiently process the video stream.
It was shown that augmentation of the input image by rotations and flips improves SR~\cite{timofte2016seven}. Following that idea, we created eight augmented input frames. The augmentation is implemented in the same way as during training, and use arbitrary flips and rotation. We used augmented frames to obtain eight predictions. These predictions were transformed back to test image reference frame, and the final SR image is a median image of the eight predicted SR images.
\subsection{pCLE-specific zero-shot training data generation}
\label{downscaling}

\subsubsection*{Pseudo-HR generation} 
The standard pCLE reconstruction algorithm relies on a Delaunay triangulation to linearly interpolate noisy fibre signals onto an oversampled Cartesian grid. 
% The fibre signals are located on the three triangle corners and form a single triangle. The pixels located inside a triangle are interpolated on the image from the corner fibre signals, effectively resulting in 
The oversampling ratio is chosen so as to minimise information loss stemming from the reconstruction, leading to an average of $7$ pixels per fibre~\cite{LeGoualher2004}.
A bundle built with 25k fibres would thus capture 25k unique fibre signals, and those are used to reconstruct a Cartesian image with around 175k pixels (excluding any black borders). Consequently, the pixel information rate is only 1/7 on the oversampled original pCLE reconstruction.

The generation of a pseudo-HR image from the input image is depicted in \figref{fig:down} in the red box.
This stage is designed to reduce redundant pixels generated by the standard pCLE reconstruction,
while allowing for some data loss, and to reduce the distortion of the pCLE textured noise pattern.
We reduce the redundancy of information in the oversampled input image by reconstructing the sparse pCLE input data on a grid $n$ times smaller, in terms of a total number of pixels, than the original one. For that, we rely on the standard pCLE reconstruction algorithm. We thus obtain an average of $n/7$ pixel per fibre in the reconstruction. Such a newly created image is referred to as pseudo-HR image as it displays much less information redundancy at the reconstructed pixel level. Thanks to reconstructing the pseudo-HR on a twice smaller grid than the original pCLE image, we create more compact information that provides high-frequency information by making the pixel information rate 4/7. This highly pack information serves as estimate of the HR that has higher density of information than the original test image. Another beneficial property coming from the reconstruction of the pseudo-HR image is that it is does not include the typical pCLE noise pattern. Thanks to that, the noise in the image does not adversely impact its information content, and potentially paves the way for better SR to be obtained.

\subsubsection*{Voronoi-based downscaling}

\begin{figure*}[!htb]
\centering
    
    \includegraphics[scale=0.7]{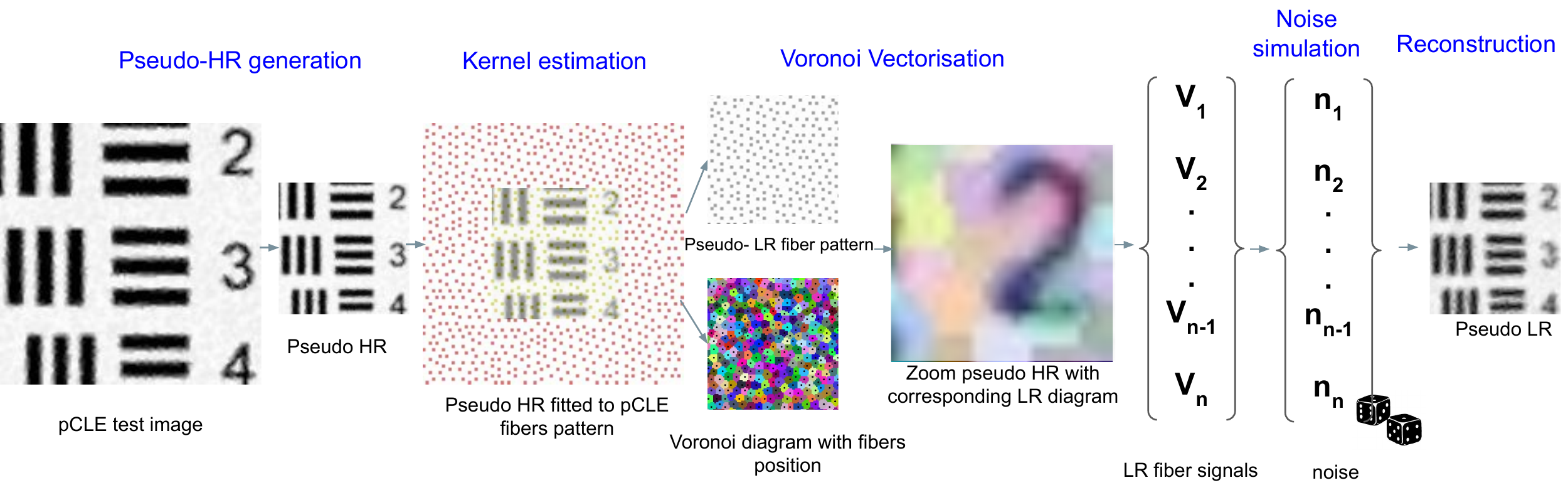}
\caption{Voronoi downscaling: A test pCLE image is reconstructed as a pseudo-HR; A pCLE fibre pattern is fitted to a pseudo-HR space to estimate kernel as a pseudo-LR Voronoi diagram; The averaged signals retrieved from the diagram by a Voronoi vectorisation are reconstructed to form a pseudo-LR pCLE.}
\label{fig:down}
\end{figure*}

A well-known limitation of the bicubic downscaling kernel when applied to natural images is its inaccurate approximation of the real LR creation process~\cite{Timofte_2017_CVPR_Workshops}.
This discrepancy is even higher in the case of pCLE given the irregular fibre-induced sampling and reconstruction characteristics of the device. Hence, a better downscaling approach is needed in our case. To generate LR pCLE images, we designed a novel downscaling procedure based on the pattern of fibre and reference pCLE reconstruction method. The downscaling procedure consists of several steps: kernel estimation, acquisition of LR signals with a Voronoi vectorisation, noise generation, and reconstruction of the LR image as illustrated in~\figref{fig:down}.

%The geometrical distribution of the fibre signals serves for construction of a Voronoi diagram used for downscaling

\textit{Kernel estimation.} In the first step, we use the pseudo-HR image to construct the pseudo-LR fibre pattern used to create a kernel in the process of Voronoi downscaling. To achieve that, we retrieve the original pCLE fibre pattern from the input image metadata. To construct the pseudo-LR fibre pattern, we associated the pseudo-HR image with the original fibre pattern. The fibres in the original pattern are distributed in a pseudo-regular pattern (quasi-hexagonal). Thus, pseudo-HR's exact position within a fibre pattern space does not play a significant role, as the characteristic distribution of fibres is preserved within a bundle. As the original pattern has a circular shape, the most straightforward implementation is to fit the pseudo-HR area to the centre of the pattern to avoid edge problems. As depicted in \figref{fig:down}, the pseudo-HR image spans a smaller area than the fibre bundle of the test image does. Fibres outside the pseudo-HR space are discarded (marked red in \figref{fig:down}). Consequently, the geometrical distribution of fibres remains unchanged, but the density of fibres per structure in the image decreases. Thanks to the fact that we reuse the fibre pattern from the input image, we ensure that we preserve the signal acquisition's nature through the fibre bundle, including typical fibre signals and their geometrical position.

\textit{Voronoi vectorisation.} We use the new fibre pattern to simulate signal loss for the acquisition of the LR signals. Every fibre signal contributes to the reconstruction of the neighbouring pixels. This contribution is defined by the Delaunay triangulation created using the position of fibres in the fibre pattern. To acquire the pseudo-LR signal, we use the pseudo-LR fibre pattern to find each fibre position. The region of influence of each fibre is given as the space covered by the corresponding cell in the Voronoi diagram, which is the dual of the Delaunay triangulation.
\par
Formally, let $X={f_1\dots f_n}$ be a set of $n$ fibres' position within fibre pattern in \textbf{R}$^d$. The Voronoi diagram generated by $X$ is the partition of the fibre pattern into $n$ convex cells, the Voronoi cells, $V_{i}$, where each $V_{i}$ contains all points of \textbf{R}$^d$ closer to $f_{i}$ than to any other point:
\begin{equation}
 V_i=\{x:\forall{j}\neq{i},d(x,f_i)\leqslant{d(x,f_j)}\},
\end{equation}
where $d(x,y)$ is the Euclidean distance between $x$ and $y$.
As presented in \figref{fig:down}, we construct the diagram in such a way that each cell contains one fibre in its centre.
Each cell in the Voronoi diagram represents the fibre's FoV and its contribution to the image pixels. Effectively, the signal in the cell $V_{i}$ is discrete and includes $m$ pixels $p$. We estimate the LR signal in each cell $LR_{i}$ by averaging the pseudo-HR $\widehat{HR}_s$ signal covered by each fibre region of influence:
\begin{equation}
    LR_{i} = \frac{1}{m}\sum_{p=1}^{m}\widehat{HR}_s(p),\; p \in V_i,
    \end{equation}
We create a vector of the pseudo-LR signal \mbox{$\overrightarrow{LR}=\{LR_i,...,LR_n\}$} of the length $n$ which equals the number of fibres in the pseudo-LR pattern. 

%These Voronoi cells act as an estimate of Voronoi kernels. 
%Pseudo-LR images emulate signal loss, which is a corollary of the use of fewer fibres for generating a pseudo-LR image in comparison to the respective test image. Because of the fewer fibres in pseudo-LR, each Voronoi cell covers relatively more informative pixels than it would in the test image space.
%Based on that, we compute LR fibre signals by assigning the average pixel value in each cell to its respective fibre. 

%We run ablation study to test importance of noise in our pipeline. 
%This allows us evaluate the effect of noise on SR reconstructions from models trained with and without simulating noise.

\textit{Noise generation.}
%\label{noise}
In the seminal ZSSR work~\cite{shocher2018zero}, the authors
studied the application of their pipeline to real, noisy images. They noticed that SR training may benefit from adding extra small Gaussian noise to pseudo-LR images. However, while Gaussian noise can be considered appropriate for natural images, this does not translate directly to our case. pCLE images are characterised by a specific type of noise produced due to the optical limitation of the system, and specific image reconstruction algorithm, which is very different to standard digital camera.
The pCLE noise has distinct texture that stems from the fact that the current reconstruction algorithm interpolates noisy fibre signals along with the tissue signal onto the Cartesian grid affecting several pixels. Depending on the experimental setup, we can also add noise to the LR signal prior to pseudo-LR image reconstruction as described below. 
Based on our previous experiences with simulating synthetic pCLE images~\cite{ravi2018effective,szczotka2020learning,ravi2019adversarial}, in order to generate realistic LR, we used the pCLE noise model suggested in~\cite{LeGoualher2004}. The authors defined two types of noise: acquisition noise that can be modelled as additive noise; and calibration imperfection that is modelled by multiplicative noise. Similarly to our previous works, we simulated both noise types by sampling them from a Gaussian distribution with a mean of zero. In contrast to our previous approach, we simulate varying levels of noise in each augmented pseudo-LR frame. We achieved that by drawing noise from Gaussian distributions with a different standard deviation value per each frame. We set sigmas $\sigma_{i}$ as $\sigma_{1}+c_{1}$ for additive noise and $\sigma_{2}+c_{2}$ for multiplicative noise, where $\sigma_{i}$ is chosen empirically and set $\sigma_{1}$ to 0.03 and $\sigma_{2}$ to 0.05, $c_{i}$ is a random number between [-0.025, 0.025] drawn from uniform distribution for each frame independently. The range was chosen empirically by visually inspecting noise at the images. 
The noise is added directly to the pseudo-LR fibre signals, before interpolation-based reconstruction. 

\par
\textit{Reconstruction.}
Finally, we reconstruct pseudo-LR with the \say{gold standard} reconstruction method based on the Delaney triangulation and interpolation. We use pseudo-LR fibre pattern to construct triangulation and acquired pseudo-LR signals with or without noise to interpolate them onto a Cartesian image of the same size as the pseudo-HR image.

\section{Experiments and Results}
\label{results}
We designed independent experiments on two data sets: synthetic and original images.
We study the application of zero-shot learning to the pCLE SR task using synthetic data. We took special care to simulate realistic endomicroscopy data to ensure that the domain gap with real pCLE data is reduced as much as possible. The experiments performed on synthetic data are used to inform the design of the experiments for real pCLE.
The best performing methodology on synthetic data is indeed evaluated on the original data set through human rating.

\subsection{Experiments on synthetic images}  
\label{synthetic_ex}
We compare super-resolved images obtained on our synthetic test set with the corresponding synthetic HR pCLE. As the baseline method, we used the reference reconstruction method currently used in the clinic, which amounts to the identity transformation in case no synthetic noise is added to the simulated LR input image. For comparison, we use IQA reference-based metrics: PSNR, SSIM, VGG loss, L1 loss, LIPIPS and Gradient Magnitude Similarity Deviation (GMSD)~\cite{xue2013gradient}. 

Based on the comparison of the super-resolved images with synthetic HR data, which serves as ground truth, we test whether there is a difference between compared video sequences on a frame-by-frame basis.

Our aim was to test the following research hypotheses (\textbf{H$\boldsymbol{i}$}):
\begin{description}
    \item [\hypotdef{voronoi}] Voronoi downscaling leads to better pCLE ZSSR than bicubic downscaling
    \item [\hypotdef{denoise}] pCLE ZSSR benefits from joint training of both the SR and the denoising tasks.
    \item [\hypotdef{video}] ZSSR benefits from expanding the training dataset from a single frame to several video frames.
    \item [\hypotdef{improve}] pCLE ZSSR performs better in comparison to SISR  and state-of-the-art DCNN
\end{description}

\begin{table*}[!htb]
\centering
\caption{Image Quality Assessment for synthetic experiments. We run ablation study to investigate the performance of models employing either Voronoi (2, 4) or Bicubic kernel (3, 5) for simulated noise or noise-free data. We compare Video ZSSR with Voronoi kernel (6) against SISR trained on synthetic data (7), state-of-the-art DCNN (8)~\cite{zhang2017beyond}, baseline interpolation reference method (2).\label{tab:synthetic}}
\begin{tabular}{|c|c|c|c|c|c|c|c|c|c|}
\hline
model &  baseline  & \multicolumn{5}{|c|}{ ZSSR } & SISR & DCNN~\cite{zhang2017beyond}\\
\hline
training &  interpolation & \multicolumn{2}{|c|}{ noise-free frame } & \multicolumn{2}{|c|}{ noisy frame } & noisy video  & simulation &  pre-trained \\
\hline
kernel &  LR (1)  & Voronoi (2) & Cartesian (3)  & Voronoi (4)  & Cartesian (5) &  Voronoi (6)  &  Voronoi (7)  &  Cartesian (8) \\
\hline
PSNR & 27.99 $\pm$ 1.08 & 28.86 $\pm$ 1.08 & 28.27 $\pm$ 0.99 & 30.16 $\pm$ 1.22 & 28.23 $\pm$ 0.93 & 30.67 $\pm$ 1.27 & 30.99 $\pm$ 1.29 & 28.04 $\pm$ 1.08 \\
\hline
SSIM & 0.851 $\pm$ 0.020 & 0.880 $\pm$ 0.017 & 0.862 $\pm$ 0.014 & 0.878 $\pm$ 0.015 & 0.849 $\pm$ 0.021 & 0.890 $\pm$ 0.014 & 0.902 $\pm$ 0.012 & 0.852 $\pm$ 0.020 \\\hline
LIPIPS & 0.781 $\pm$ 0.017 & 0.806 $\pm$ 0.015 & 0.785 $\pm$ 0.012 & 0.806 $\pm$ 0.014 & 0.777 $\pm$ 0.018 & 0.817 $\pm$ 0.012 & 0.816 $\pm$ 0.014 & 0.782 $\pm$ 0.017 \\\hline
GMSD & 0.940 $\pm$ 0.006 & 0.954 $\pm$ 0.007 & 0.953 $\pm$ 0.004 & 0.955 $\pm$ 0.004 & 0.943 $\pm$ 0.006 & 0.960 $\pm$ 0.004 & 0.961 $\pm$ 0.004 & 0.940 $\pm$ 0.006 \\\hline
L1 loss & 0.973 $\pm$ 0.003 & 0.975 $\pm$ 0.003 & 0.974 $\pm$ 0.003 & 0.980 $\pm$ 0.003 & 0.973 $\pm$ 0.003 & 0.981 $\pm$ 0.003 & 0.982 $\pm$ 0.003 & 0.973 $\pm$ 0.003 \\\hline
VGG & 2.62 $\pm$ 0.19 & 2.39 $\pm$ 0.17 & 2.70 $\pm$ 0.14 & 2.37 $\pm$ 0.16 & 2.69 $\pm$ 0.20 & 2.24 $\pm$ 0.14 & 2.25 $\pm$ 0.16 & 2.60 $\pm$ 0.19 \\\hline

\end{tabular}
\end{table*}
For each research hypothesis above, we constructed a corresponding statistical null hypothesis and analysed the statistical significance of the IQA improvement with the two-sided student t-test using a significance level of $p=0.05$.
We evaluate the proposed research hypotheses with predictions generated by several models. The quantitative results from our IQA on synthetic pCLE is presented in~\tabref{tab:synthetic}. 
Here, we provide an experimental evaluation of the results per hypothesis for our ablation study.

\subsubsection*{Study of the downscaling kernel choice (\hypotref{voronoi})}
In our ablation study, we compare super-resolved images generated by models trained with our proposed Voronoi kernel against models trained with a baseline bicubic kernel. We train models for both downscaling kernels independently. We used four models, marked in Table~\ref{tab:synthetic} as 2-5, which are also trained to study the effect of noise on the pipeline.
\par
As described in more detail in \secref{downscaling}, the pseudo-HR image is obtained by reconstructing the input image on a grid twice times smaller than the original grid. Based on visual inspections of the images, we found that this downscaling rate is optimal for creating a smaller image without lose of information. This new image is used in place of the input image during a training phase serving as an HR image. 
\par
The pseudo-HR is also a part of the pCLE downscaling procedure described in \secref{downscaling} and marked by \textbf{D} in \figref{fig:pipe}. The protocol for downscaling with the bicubic kernel is designed as follows. First, we downscale the pseudo-HR image using the native TensorFlow \texttt{tf.image.resize} function with the bicubic kernel, anti-aliasing and an empirically chosen scale value of 3. Our ZSSR architecture is designed to use the LR and HR images with the same size and does therefore not embed any upsampling layer. To comply with our pipeline, we upsample the downscaled intermediate image with the bicubic upsampling to go back to the size of pseudo-HR. Since the Cartesian downscaling does not use fibre patterns, it is impossible to simulate typical pCLE noise with it. Nonetheless, we add multiplicative and additive noise pixel-wise on the Cartesian LR images for training model 5 only.
\par
The results in \tabref{tab:synthetic} demonstrate the suitability of the Voronoi kernel in ZSSR. Utilising Voronoi downscaling yields consistently better results
than bicubic downscaling for all metrics. We run two-sided paired t-test for all frames inferred by models 2, 3 and 4, 5 which confirm that there is a statistically significant improvement in image quality of the reconstructed SR images from models exploiting Voronoi downscaling. From the observer perspective, videos enhanced by models taking advantage of Voronoi downscaling are visibly sharper than those that use bicubic downscaling.
Therefore we confirm that models trained with images downscaled with Voronoi kernel yield better reconstructions than ones trained with Cartesian downscaling.
\par
\subsubsection*{Study of the impact of simulated noise on ZSSR (\hypotref{denoise})} We trained our ZSSR pipeline with two types of data: noise-free and noisy pseudo-LR. To test the influence of noise on the image quality of pCLE, during every training iteration, we augmented the LR frame by simulating noise. Because of ZSSR limits training to one data sample, it is essential to augment the training dataset, ensuring variability of transformation. Thus we draw a new random noise signal each time weights are updated.
This makes the network capable of capturing not only one noise pattern characteristic for the test image, but also the random nature of the noise. The experiments were repeated for each downscaling kernel. The quantitative scores are shown in \tabref{tab:synthetic} for models 2-5.
%The visual results are depicted in Figure~\ref{fig:syn_rec}. 

It is unsurprising to find that the models trained on noise-free images (2,3) produce models unable to distinguish between true signal and noise on the test pCLE images, while models which account for noise (4,5) generalise towards denoising.
One of the key findings is that for pCLE data image quality improvement can be implemented for both super-resolving and denoising capabilities in one model. 
This holds true for both types of downscaling kernels, yet models trained with Voronoi kernel tailored to the specific pCLE acquisition noise yield results with significantly higher PSNR of 2.0 dB. We attribute it to the fact that Cartesian noise does not represent typical pCLE patterns well. Therefore, this is supporting evidence that our physically inspired downscaling Voronoi kernel accounting for pCLE noise fits the pCLE ZSSR task better. Voronoi downscaling models noise patterns by interpolating it during the reconstruction of the image from the noisy fibre signals, which we demonstrate to be a crucial element of pCLE denoising.
Lastly, it is important to note that the models trained only for SR task not taking noise into consideration generate higher SSIM, than models which perform SR and denoising together. Although SSIM scores are slightly lower for the SSIM on a combined SR and denoising task, we attribute that difference to the fact that SSIM is a robust metric in the presence of the noisy signal. SSIM does not measure the difference in noise between images precisely enough, since it is more suited as a structural metric, and PSNR is a better metric in this case. There is a minimal difference for GMSD, LIPIPS and VGG in favour of models trained with noisy data. This small difference like in case of SSIM may be attributed to the fact that these metrics are not designed to measure the impact of noise only, but rather complex structural similarity. The L1 loss behaves similarly to PSNR giving better result for models aware of the noise. 
Informal visual inspection of the images aligns with quantitative results. It is noticeable that images with reduced noise are characterised by higher quality.
\par
\subsubsection*{Study of the training set size extended to multiple-frames training in video ZSSR (\hypotref{video})}
The previous experiments were limited to use of the only one (first) frame from the video sequence. We observe that several consecutive frames show similar content.
Frames are correlated and characterised by small information entropy. These frames altogether are nonetheless more informative than one frame, and can be used to build a more robust training dataset. Importantly, the video is acquired with the same fibre bundle, and from the same tissue and patient, thus one downscaling kernel can be used for all frames in the video. Based on this observation, we expand our training dataset from 1 single frame to the first 10\% of the frames from the video sequence, which is around 6 frames for each video.

The results in \tabref{tab:synthetic} obtained from the model trained with several frames only slightly outperform the models trained with one frame with statistical significance. The improvement for PSNR is small; only 0.5dB when compared with the model trained on one image and Voronoi kernel with noise simulation. It may suggest that it is sufficient to train the denoising model with one frame, which is augmented every iteration with newly drawn noise. Although the increase in the number of frames gives only a small increase in PSNR and L1, it significantly benefits SSIM, LIPIPS, VGG and GMSD. Since SSIM, LIPIPS, VGG and GMSD measures the structural similarity between images, we can conclude that the extended dataset benefits SR more than the denoising task.
On the other hand, there is only a modest improvement visible the naked eye during informal visual inspections of videos. 
These findings reinforce the general belief that bigger datasets help in the training of neural networks, which is also valid for zero-shot learning.

\subsubsection*{Study of performance of training type -- Supervised SISR and pre-trained state-of-the-art DCNN vs unsupervised ZSSR (\hypotref{improve})}
In order to provide a comparison of unsupervised zero-shot learning to the supervised single image SR, we also trained
the same network architecture presented in Figure~\ref{fig:pipe} using a SISR approach. Similarly to our previous study~\cite{szczotka2020learning}, we used the synthetic test and train dataset, and trained the residual mapping between synthetic pairs of HR-LR images. We also compared ZSSR to state-of-the-art network for joint denoising and SR called DCNN with pre-trained model 'dncnn3'\footnote{ Available at  \url{https://github.com/cszn/KAIR}}. Those models are used to generate a prediction for a synthetic test set.

%In line with the findings of these experiments, we 
We
report in \tabref{tab:synthetic} that
ZSSR statistically does not outperform SISR.
%Quite opposite 
As a matter of fact,
supervised SISR performs significantly better. However, our findings are not surprising since it is expected that models trained in a supervised manner perform better than the one trained as unsupervised as in the case of ZSSR.
Additionally, the informal visual inspection does not capture any significant differences between videos, which suggest that both models are trained towards the same/similar solution. This, in turn, provides more confidence to unsupervised ZSSR in the reconstruction of pCLE images. It also highlights that kernel choice may be a powerful driver towards converging models to the optimal solution, and data size while still important may play a smaller role. On the other hand, ZSSR outperforms DCNN model by a large margin. It is related to both the fact that DCNN is pre-trained with natural images for denoising the Gaussian noise and uses bicubic kernel as a downscaling process. We have shown that both of these elements are not suitable for pCLE, in \textbf{H1} and \textbf{H2}, as it has very distinct noise type and Voronoi kernel mimics downscaling better than the a bicubic kernel.

%To sum up our investigation of the ZSSR on synthetic data, we identify that the best performing approach is to use video ZSSR for the joined task of SR and denoising with use of Voronoi downscaling. Both informal visual inspection and quantitative analysis confirmed that these models yield SR images with the best quality. 

\subsection{Experiments on original images}
In our experiments on synthetic data, we demonstrated that most significant improvement in the image-quality of pCLE is achieved by the supervised SISR and the unsupervised video ZSSR models. In this section, we describe how we translated these two models to our original data, and evaluate the quality of obtained reconstructions with reference to each other and the baseline interpolation method by subjective assessment of the image quality performed as a user study.

Thanks to the reference-free ZS training scheme, we can train SR models on original pCLE videos, not relying on (unavailable) real HR images. We trained seven independent ZSSR models, one per test video selected from our original dataset described in \secref{data}.
%The good performance of ZSSR in application to pCLE domain is attributed to the pCLE specific implementation including customisation with the crafted downscaling kernel, simulation of specific pCLE noise, and training on a test video sequence itself. Therefore, we
We
designed Voronoi kernels specifically to account for the acquisition of each original video with its unique fibre pattern. We also simulated pCLE-like noise with sigmas $\sigma_1$ and $\sigma_2$ set as 0.1 for additive noise and 0.5 for  multiplicative  noise. The sigmas were chosen based on informal visual inspection of images.
%We attribute stronger noise patterns to the fact that scaling original images to smaller fibre bundle is more prone to generating stronger noise patterns.
For the training, we used the first 10\% of all frames in each video, which have around 60 frames in total. This number of frames allows to create a representative sample of the video and allows for fast training. Once models converge, we interfere results for all the frames in the test video on which the model was trained creating SR video reconstruction.
\begin{figure}
\centering
  \includegraphics[scale=0.47]{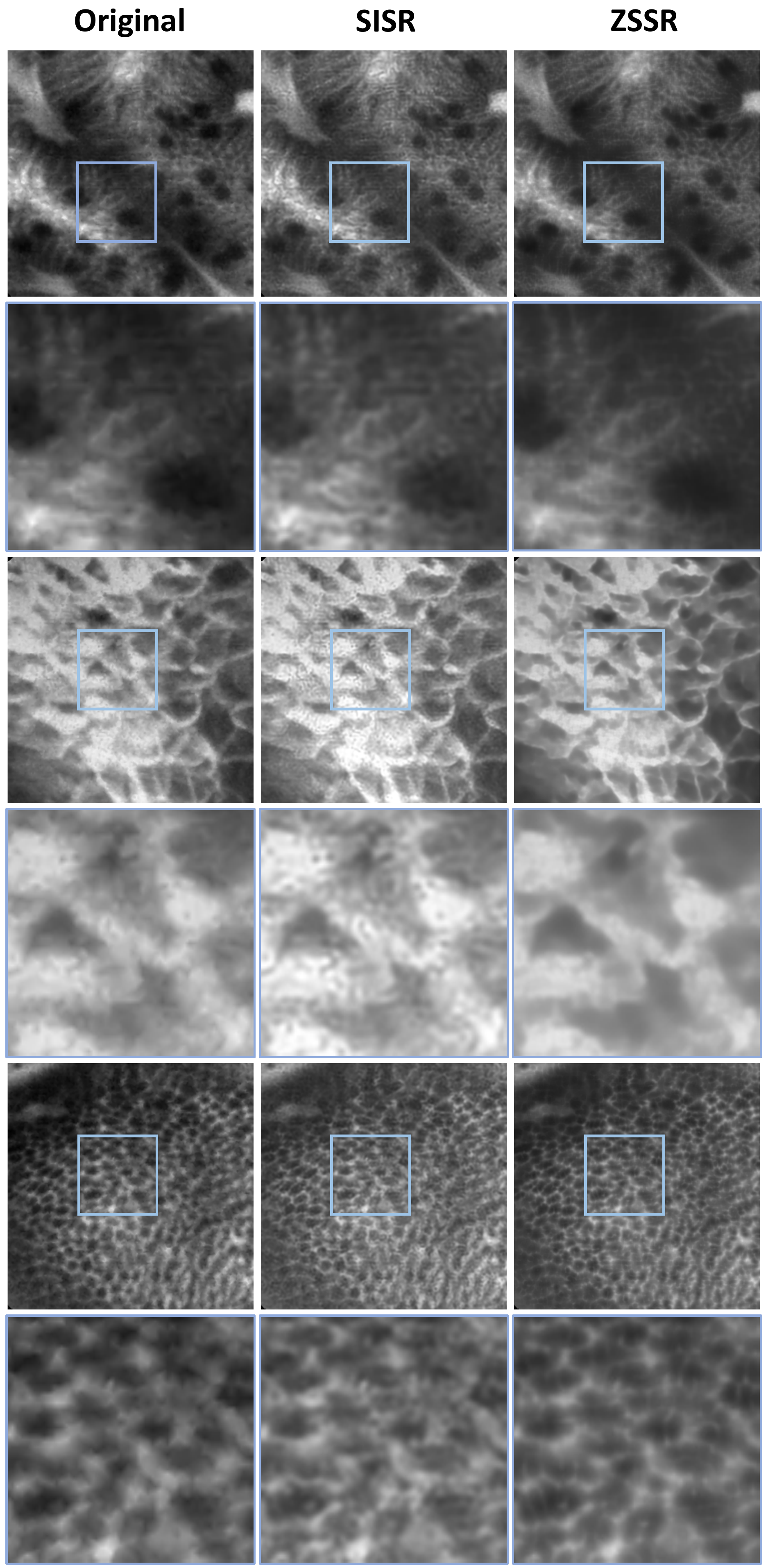}
  \caption{The results for two different tissues for the baseline, SISR and video ZSSR (starting from the left). Each odd row present patch from an original pCLE image; each even row presents the zoomed square marked with blue on the image in the row above it.}
  
  \label{fig:resultim}
\end{figure}
While ZSSR has a blind training protocol that allows creating video-specific models directly from original data,
SISR cannot be easily retrained without ground truth HR pCLE images. Nonetheless, it was shown that original pCLE reconstruction benefits form models trained on synthetic data~\cite{ravi2018effective}. Thus, we re-use the SISR model trained on synthetic data presented in \secref{synthetic_ex}
%\textit{Experiments on synthetic images}
and use it to infer results for the original test videos. Thanks to the realistic simulation of the synthetic training dataset, we expect that this SISR model performance translates well for a task of improving the quality of original data. The example results are depicted at the \figref{fig:resultim}.
\par  
In the absence of original ground truth HR images as reference, well-established reference-based metrics are ill-suited to assess the quality of the super-resolved images obtained from this data. To evaluate our proposed methodology in application to real pCLE videos, we asked observers to compare images reconstructed from three methods:  ZSSR, SISR and baseline. The preferences of the observers indicate an improvement in the pCLE image quality. We design our user study to test the following research hypotheses (\textbf{H$\boldsymbol{i}$}):
\begin{description}
    \item [\hypotdef{zs_baseline}] video ZSSR with simulated noise yields super-resolved pCLE images, which are preferable to the baseline reconstruction by both experts and non-experts.
    \item [\hypotdef{sisr_baseline}] SISR yields pCLE super-resolved image which is preferable to the baseline reconstruction by both experts and non-experts.
    \item [\hypotdef{zs_sisr}] video ZSSR with simulated noise outperforms SISR in improving the image quality of the reconstructed pCLE by the judgement of both groups of raters.
\end{description}

\begin{subsection}{User study}
Among the most common user studies for classifying image quality such as single‐stimulus, double‐stimulus, and similarity judgements, it was shown that forced‐choice pairwise comparison is the most accurate and  time-efficient~\cite{mantiuk2012comparison}. The most consistent results are generated by a cognitively easy approach.  In the forced‐choice approach, observers need to compare only two images at a time and make a quick binary decision without any rating scales. Based on the suitability of a forced-choice approach, we designed an image quality assessment survey based on two-alternative forced-choice (2AFC) paradigm to test hypothesis 
%\hypotref{voronoi}-\hypotref{video}.
\hypotref{zs_baseline}-\hypotref{zs_sisr}.

\subsubsection*{Survey structure}
We compare three methods by directly inferred pair comparisons: baseline vs SISR, baseline vs ZSSR, and SISR vs ZSSR. It is essential to consider that pCLE reconstructions are similar, and the cyclic relations may occur. To avoid any error coming from such cyclic relationships, we decided to test all possible combinations giving three direct comparisons.

The images used in the survey were extracted from all test videos for each of methods: SISR, ZSSR and baseline. For each comparison, we extracted 12 random frames from each video. Frames are used only once. The same frame is never re-used in another comparison. Thanks to extracting multiple frames from each video, we can assess results on both frame basis and cumulative video basis. We ensure that each video comes from a different tissue, patient and fibre assuring high variability of data dynamic range among real clinical cases.

Each question tests one of the three comparisons. There are 252 questions in the entire survey, 84 questions for each comparison pair. All questions were randomised to avoid correlations between questions such as same video, frame or model tested in adjacent questions. For each question, two images from two different methods are displayed in a random order, and with different randomisation.

\subsubsection*{Observers}
The image quality of the pCLE reconstructions is assessed by the subjective perception of image quality of expert and non-expert observers. 
All recruited in our study experts have more than two years of experience working with pCLE images. Non-expert users did not have exposure to pCLE images prior to the survey. 
\begin{table}[tbh!]
\centering
 \caption{Summary of preferences for Image Quality Assessment survey for both  experts and nonexperts. There are three tests: Baseline vs. ZSSR (1); baseline vs. SISR (2); and SISR vs. ZSSR (3). The preferences are given as the percentage (\%) of frames chosen by all observers in each comparison.\label{tab:survey}}
\begin{adjustbox}{width=\columnwidth}
\begin{tabular}{|c|c|c|c|c|c|c|}
\hline
& \multicolumn{2}{|c}{ Test 1 } & \multicolumn{2}{|c|}{ Test 2 } & \multicolumn{2}{c|}{ Test 3 } \\\hline
model & BASELINE & ZSSR & BASELINE & SISR & SISR & ZSSR \\\hline
experts& 23 & 77 & 64 & 36 & 25 & 75 \\\hline
nonexperts& 36 & 64 & 25 & 75 & 62 & 38 \\\hline
\end{tabular}
\end{adjustbox}
\end{table}

\subsubsection*{Experimental procedure} 
The observers received the survey as a website. % a link to the survey implemented as a %surveying
%web application\footnote{We relied on \url{https://www.surveylegend.com/}}.
On the first page, the display conditions of the survey were given to the observers. The observers were asked to run a survey on a computer screen size bigger than 13'' in full window mode, 
to ensure visualisation conditions mimicking a clinical set-up.
%as this simulates a clinical set-up the best.
Second, they were instructed how to use the survey tool for more detail comparison with the ability to zoom in the images. Lastly, users were shown a sample question. Since the task of comparing two images is intuitive and relatively easy for human observers, it does not require extensive training before starting the survey.

The observers were shown two pCLE reconstructions arising from two of the tested methods next to each other. They were asked to select a higher quality image. Users were instructed to assess images in less than 15 seconds per question, but no time restriction was imposed by the survey tool. The observers spent 26 minutes on average to fill in the survey.
%Observers could use zoom to compare images in detail.
They had to answer every question in the survey and were not allowed to come back to a previous question.

Although categorical data can be ranked via direct vote count, additional care is needed to perform statistical analysis.
Hypothesis testing of 2AFC requires 
%non-trivial data analysis which requires 
modelling a quality scale based solely on the comparison ranks.
Among the most popular methods for pairwise comparisons in image quality is the Thurstone-Mosteller model~\cite{handley2001comparative}. Particularly, we used the Thurstone Case V scaling method with outlier analysis and statistical testing\footnote{Available at \url{https://github.com/mantiuk/pwcmp}} to analyse the observer preferences.
%to test our hypothesis.

The responses to the survey are summarised in \tabref{tab:survey}. There were 10 experts and 40 non-experts taking part in the survey. Since expertise plays a key role in understanding images, we stratify the analysis by the group.
%The data from the survey suggest that:

\subsubsection*{Findings relating to \hypotref{zs_baseline}} %
The results for both observer groups show broadly similar traits. The experts show clear preference by giving 77\% of votes towards ZSSR reconstructions over baseline solution with the statistical significance confirmed by fitting a Thurstone-Mosteller model.
The preference towards ZSSR is also seen among non-expert users, yet results vary between videos, and statistical significance was not achieved for their cumulative choice.

\subsubsection*{Findings relating to \hypotref{sisr_baseline}}
An interesting pattern emerges when examining the preferences for SISR. It is very apparent that experts prefer the baseline method (64\% of votes) over the learning-based SISR approach. On the other hand, non-expert observers showed polar behaviour and chose the SISR model as their favourite (75\% of votes).
Both votings are shown statistically significant in our analysis.  This extreme difference may be attributed to the understanding of the pCLE domain and experts' ability to reject unnaturally enhanced images with possible non-informative domain-specific artefacts. 

\subsubsection*{Findings relating to \hypotref{zs_sisr}}
Experts prefer ZSSR (75\% of votes) over SISR, and non-experts choose SISR instead (62\% of votes). Consequently with previous results, we observe the same pattern throughout the survey, which confirms the cyclic relationship for the performance of the reconstruction method. This confirms with statistical significance that both tested groups behaved consistently.

The findings indicate that both groups find the quality of learning-based super-resolved image better than baseline interpolation methods, although observers exposed varying opinion of which model is better. The detailed statistical analysis with visualisation and tested videos with their SR versions are included in the supplementary material. 
\par
In addition to the user study outcomes, we provide some qualitative assessment of the results. Both DL methods restore images with higher contrast and enhanced details. Also, the triangulation artefact, which is seen on original images reconstructed with the baseline method, is removed by both DL methods. The main distinction between ZSSR and SISR reconstructions is the denoising power of ZSSR. SISR reconstruction contains characteristic pCLE noise that is even enhanced compared to original images, while ZSSR reconstructions are significantly de-noised. Also, ZSSR reconstruction reveals small structures which are not distinguishable from noise. Overall, we find enhanced images easier to interpret as the biological structures are more apparent when noise and artefacts are removed. Those pixel-level changes are an essential clue for the clinician allowing differentiating cellular structures. We may see the clinical impact of that changes as a more robust source of information for the diagnosis than previously available with standard reconstruction methods.
\end{subsection}

\section{Conclusions}
We proposed a novel method for improving the quality of endomicroscopy. Our pipeline combines Zero-Shot learning which encapsulates a downscaling kernel tailored to the acquisition physics by incorporating fibre bundle geometry and noise simulation. 
Zero-Shot Super-Resolution (ZSSR) is a patient-specific training approach that considers all available information and parameters, such as fibre bundle type and its unique fibre pattern, tissue type, and clinical case.
It also makes ZSSR immune to external data such as another fibre configuration, tissue type, or clinical case, which could potentially be misleading.
The specificity of our method is not the only advantage. Its flexibility is deemed of high interest since the method can be trained with very little data in comparison to state-of-the-art supervised approaches. Limited necessary computational resources make it easier to implement in a real-world clinical set-up, for example as a reconstruction algorithm using the first few frames for calibration. The quantitative and qualitative image assessment confirms the superiority of our ZSSR reconstruction method over the baseline interpolation-based reconstruction algorithm. The baseline method does not use any prior information, but just naive interpolation to reconstruct images. In comparison, DL-based reconstructions use multiple kernels trained on prior information provided by the pCLE dataset. These kernels, which are trained to reconstruct HR from LR images, allow for more informed reconstructions of the images that enhance their quality. ZSSR serves as a more effective alternative to SISR, which is itself restricted in applicability to real pCLE domain due to the lack of ground truth HR pCLE images.

\bibliographystyle{IEEEtran}
\bibliography{zssr.bib}
\end{document}